# Enhancing the performance of a safe controller via supervised learning for truck lateral control

Yuxiao Chen[ab], Ayonga Hereid[c], Huei Peng[b] and Jessy Grizzle[c], *Fellow, IEEE*

*Abstract*—Correct-by-construction techniques, such as control barrier functions (CBFs), can be used to guarantee closed-loop safety by acting as a supervisor of an existing or legacy controller. However, supervisory-control intervention typically compromises the performance of the closed-loop system. On the other hand, machine learning has been used to synthesize controllers that inherit good properties from a training dataset, though safety is typically not guaranteed due to the difficulty of analyzing the associated neural network. In this paper, supervised learning is combined with CBFs to synthesize controllers that enjoy good performance with provable safety. A training set is generated by trajectory optimization that incorporates the CBF constraint for an interesting range of initial conditions of the truck model. A control policy is obtained via supervised learning that maps a feature representing the initial conditions to a parameterized desired trajectory. The learning-based controller is used as the performance controller and a CBF-based supervisory controller guarantees safety. A case study of lane keeping for articulated trucks shows that the controller trained by supervised learning inherits the good performance of the training set and rarely requires intervention by the CBF supervisor.

*Nomenclature*

$\mathbb{R}$ denotes the set of real number, $\mathbb{R}^n$ denotes the n-dimensional Euclidean space, $\mathbb{R}[x]$ denotes the space of all polynomials of $x$, $\Sigma[x]$ denotes the cone of SOS polynomials, a subset of $\mathbb{R}[x]$. For a scalar function $h: \mathbb{R}^n \to \mathbb{R}$ of $x \in \mathbb{R}^n$ and a vector field $f: \mathbb{R}^n \to \mathbb{R}^n$, the Lie derivative is defined as $\mathcal{L}_f h(x) = \frac{dh}{dx} f(x)$, which is a scalar function of $x$, and $\mathcal{L}_f^n h = \mathcal{L}_f \mathcal{L}_f^{n-1} h$, with $\mathcal{L}_f^0 h = h$. $C^n$ denotes sets of functions with continuous *n-th* derivatives.

## I. INTRODUCTION

CORRECT-by-construction control synthesis has been a promising direction of research that brings formal safety guarantees to controller design. In particular, Control Barrier Functions (CBF) can be overlaid on existing controllers so as to impose closed-loop safety in a plug-and-play fashion [1, 2]. The key idea in the design of a CBF is to compute a forward invariant set that contains the safe set and excludes the danger set. The CBF can then be implemented in a supervisory control structure to guarantee safety without redesigning the performance controller, hereafter called the 'student' controller because it is being 'supervised' by the CBF.

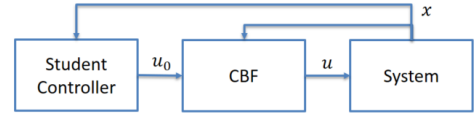

Fig. 1. Block diagram of Control Carrier Function (CBF) supervising a "student" controller.

As shown in Fig. 1, $u_0$ denotes the control input from the student controller, which can be designed with any existing method, and $u$ denotes the input signal after the intervention of the supervisory controller. If $u_0$ respects the safety constraint, then $u = u_0$; otherwise a 'minimal intervention' is applied. Depending on the form of the barrier function, the 'intervention' may be computed through quadratic programming [1, 3], mixed integer programming [4], or in other forms.

While safety is assured independently of the choice of student controller, if the student controller is not properly designed, or is designed in a way that is not compatible with the CBF, the CBF may be triggered frequently, leading to undesirable closed-loop performance. In [5], when working with a student controller for Adaptive Cruise Control (ACC) that is not properly designed, the CBF causes spikes on the input when activated. In [4], when the student controller is designed without considering obstacle avoidance, the CBF has to intervene frequently and severely to ensure obstacle avoidance. These examples, on one hand, demonstrate the power of a CBF to provide safety guarantees, but they also show there is room for improvement. If the student controller is designed in a way that takes the supervisor into account, the interventions can be reduced and the overall system's performance can be improved.

On the other hand, machine learning has been used extensively in dynamic control. Supervised learning has been used to learn a control policy with structure[6, 7], deep learning recently was used to generate end-to-end Lane Keeping (LK) policy, i.e., a mapping directly from the camera pixels to the steering input [8], and reinforcement learning can be used to generate a control policy in an 'explore and evaluate' manner

[a] This work was supported in part by the NSF under Grant CNS-1239037 and in part by the Toyota Research Institute.
[b] Yuxiao Chen and Huei Peng are with the Department of Mechanical Engineering, University of Michigan, Ann Arbor, MI, 48109 USA e-mail: {chenyx,hpeng}@umich.edu.
[c] Ayonga Hereid and Jessy Grizzle are with the Department of EECS, University of Michigan, Ann Arbor, MI, 48109 USA e-mail: {ayonga, grizzle}@umich.edu.

[9-11]. However, one major deficit of machine learning is its extreme difficulty for analysis. The number of parameters contained in a neural network can easily reach several thousand, even millions, which makes it practically impossible to analyze. Therefore, the safety of a learning-based controller should rely on other tools, such as reachable sets and barrier functions. In this sense, machine learning and CBFs complement one other.

Existing methods that combine learning with safety guarantee include reachable-set-based learning scheme that can guarantee safety for online learning of a control policy [12, 13], a barrier-function-based online learning scheme [14], and Gaussian process learning [15]. Unlike approaches that aim at guaranteeing safety with learning, such as [12-15], the method proposed in this paper separates safety from the performance. The safety guarantee is provided by a CBF, and supervised learning is used to improve the performance considering the influence of the CBF as a supervisor.

The method we propose is to perform trajectory optimization offline, generate a library consisting of trajectories with good properties, namely, stabilizing an equilibrium, attenuating disturbances, and satisfying a CBF condition. We then use supervised learning to design a student controller that inherits the properties of the trajectory library. The CBF is implemented as a supervisor of the learning-based controller, as shown in Fig. 2. Since the CBF condition is enforced in the training set, an intervention by the supervisor is rarely triggered. It should be emphasized that the safety is still guaranteed by the CBF, the supervised learning only aims to improve performance.

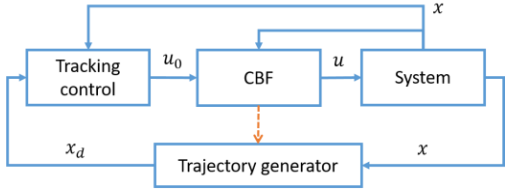

Fig. 2. Structure of the proposed supervisory control

The main contributions of this paper are the following two points. First, we propose a supervised learning based method to design a student controller that takes the CBF condition into account, is applicable to a large region of initial conditions, and rarely triggers an intervention from the supervisory controller. With supervised learning, the design of a safe student controller is transformed into the design of safe trajectories, which is much easier, as conceptually shown in Fig. 3.

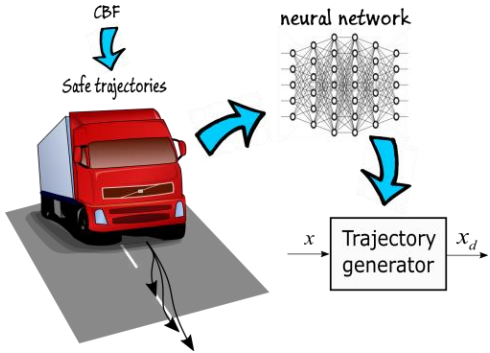

Fig. 3. Learning based trajectory generator

Second, we provide a stability and set invariance analysis of the learning-based controller under the framework of continuous hold (CH) feedback control. Applying the proposed method, we are able to provide a safety guarantee for Lane Keeping control (LK) of an articulated truck, while achieving good ride comfort.

The remainder of the paper is structured as follows. We first introduce the truck model and the feedback linearization structure in Section II. Then we present the Sum of Squares (SOS) approach for the synthesis of a CBF in Section III. Then we show the trajectory optimization process with direct collocation that incorporates the CBF condition in Section IV. The obtained trajectory library is then used to train a neural network that acts as a trajectory generator, as presented in Section IV. The trajectory generator is implemented in a Continuous Hold control structure with CBF as the supervisor on top of it, which is presented in Section V. Finally, we present the LK problem as an example in Section VI and conclude in Section VII.

## II. DYNAMIC MODEL AND VIRTUAL CONSTRAINT

In this paper, we consider a control affine nonlinear model:
$$\dot{x} = f(x) + g(x)u + g_{d1}(x)d_1 + g_{d2}(x)d_2, \\ x \in \mathbb{R}^n, u \in \mathcal{U} \subseteq \mathbb{R}, d_1 \in \mathcal{D}_1 \subseteq \mathbb{R}^{l1}, d_2 \in \mathcal{D}_2 \subseteq \mathbb{R}^{l2} \quad (1)$$

where $x$, $u$, $d_1$, and $d_2$ represent the state, input, measured disturbance, and unmeasured disturbances, respectively.

**Remark 1**: *The unmeasured disturbance $d_2$ will be countered with the feedback control. Therefore, it is assumed that $d_2 = 0$ for the following analysis of feedback linearization.*

### A. Model assumptions

The results in this paper are developed under four key assumptions:

**A-1**: *It is assumed that $d_1$ changes slowly comparing to the system dynamics. Therefore, $d_1$ is treated as constant in the following analysis.*

**A-2**: *There exists an output $z = h(x)$ for $x$ within an open subset $\mathcal{S} \in \mathbb{R}^n$, such that for all $d_1 \in \mathcal{D}_1$, $z$ has relative degree $\rho$, where the relative degree is defined as the integer such that $\forall x \in \mathcal{S}$,*
$$\mathcal{L}_g \mathcal{L}_{\bar{f}}^{i-1} h(x) = 0, i = 1, 2, ..., \rho - 1; \\ \mathcal{L}_g \mathcal{L}_{\bar{f}}^{\rho-1} h(x) \neq 0, \quad (2)$$

where
$$\bar{f}(x) = f(x) + g_{d1}(x)d_1. \quad (3)$$

**A-3**: *It is assumed that when $d_2 = 0$, for all $d_1 \in \mathcal{D}_1$, there exists a unique $\eta_u(d_1) \in \mathcal{U}$ that maintains a unique equilibrium point $x_e \in \mathbb{R}^n$ with $h(x_e) = 0$, denoted as $x_e = \eta_x(d_1)$:*
$$f(x_e) + g(x_e)\eta_u(d_1) + g_{d1}(x_e)d_1 = 0, \\ h(x_e) = 0. \quad (4)$$

Then from feedback linearization, there exists a state transformation:

$$\begin{bmatrix} \sigma \\ \xi \end{bmatrix} = \begin{bmatrix} T_1(x) \\ T_2(x) \end{bmatrix} = T(x),$$

$$\sigma \in \mathbb{R}^{n-\rho}, \xi = \begin{bmatrix} h(x) \\ \vdots \\ \mathcal{L}_f^{\rho-1} h(x) \end{bmatrix} = \begin{bmatrix} z \\ \vdots \\ z^{(\rho-1)} \end{bmatrix} \in \mathbb{R}^{\rho} \quad (5)$$

where $T$ is a bijective diffeomorphism over $\mathcal{S}$, and the transformation satisfies $\frac{\partial T_1}{\partial x} g(x) = 0$. Therefore, the dynamics of the "hidden" states $\sigma$ is represented as

$$\dot{\sigma} = \Gamma(\sigma, \xi). \quad (6)$$

In particular, $\dot{\sigma} = \Gamma(\sigma, 0)$ is the zero dynamics of the system with output $z$, and there exists a smooth surface $\mathcal{Z} \subset \mathcal{S}$ defined by $\mathcal{Z} := \{x \in \mathcal{S} | \xi = 0\}$, which is the zero dynamics manifold.

**A-4**: *We assume that the zero dynamics of the system under output $z$ is exponentially stable within $\mathcal{S}$.*

Then by Theorem 11.2.3 in [16], the following feedback linearization controller constructed from $z$ and its derivatives stabilizes the equilibrium $x_e$:

$$u = -\frac{1}{\mathcal{L}_g \mathcal{L}_f^{\rho-1} h(x)} \left[ k_0 \xi_1 + \ldots + k_{\rho-1} \xi_\rho + \mathcal{L}_f^\rho h(x) \right], \quad (7)$$

where $\{k_i\}$ is a set of exponentially stabilizing gains in the sense that the following characteristic equation

$$\lambda^\rho + k_{\rho-1} \lambda^{\rho-1} + \ldots + k_0 = 0 \quad (8)$$

has all of its roots in the open left half plane. See e.g. [17] for reference on feedback linearization and zero dynamics.

*B. Virtual constraint and tracking control*

To let the system track a desired trajectory of $z$, we use the virtual constraint method, originally developed in the robotics literature [18-20], and now appearing more widely. Suppose we want the system to track the following trajectory:

$$z = h_{des}(t), \quad (9)$$

where $h_{des}$ is a $\rho$ times continuously differentiable function. Differentiate (9) $\rho-1$ times and define the error states:

$$\begin{aligned} e_1 &= z - h_{des} \\ e_2 &= \dot{z} - \dot{h}_{des} \\ &\vdots \\ e_\rho &= z^{(\rho-1)} - h_{des}^{\rho-1} \end{aligned} \quad (10)$$

Then pick $\{k_i\}$ to be a set of stabilizing gains as described in (8), and let

$$u = -\frac{1}{\mathcal{L}_g \mathcal{L}_f^{\rho-1} h(x)} \left[ k_0 e_1 + \ldots + k_{\rho-1} e_\rho + \mathcal{L}_f^\rho h(x) \right], \quad (11)$$

When $h_{des}$ is $\rho$ times continuously differentiable and its derivatives are bounded, the feedback linearization control can locally track $h_{des}$ imposed as a virtual constraint of $z$ [21].

The benefit of the virtual constraint approach is that it gives a simple means of parameterizing the desired evolution of the vehicle. Instead of all the states, the desired trajectory is parameterized only by an output $z$ satisfying A-2 and A-4. Later, we will use trajectory optimization to determine the existence of a set of interesting trajectories that can be tracked by considering the full dynamics and the feedback structure.

*C. Tractor-semitrailer models*

In this work, we use two models: a design model and a validation model. For validation, we use TruckSim with its impressive 312 states. The literature contains a range of less detailed models that could be considered for control design, ranging from the nonlinear 37-state, physics-based model in [22], to linear models. To demonstrate the fundamental robustness of the approach followed in this paper, we base the control design on a low-complexity model for an articulated truck adapted from [22] and [23], namely a 4 DOF linear model with 8 states:

$$x = \begin{bmatrix} y & v_y & \psi & r & \psi_a & r_s & \phi & p \end{bmatrix}^T \quad (12)$$

where $y$ is the lateral deviation from the lane center to the tractor Center of Gravity (CG), $v_y$ is the lateral sideslip velocity of the tractor, $\psi$ is the heading angle of the tractor, $r$ is the yaw rate of the tractor, $\psi_a$ is the articulation angle on the fifth wheel (the joint between the tractor and semitrailer), $r_s$ is the yaw rate of the semitrailer, $\phi$ is the roll angle and $p$ is the roll rate, as shown in Fig. 4.

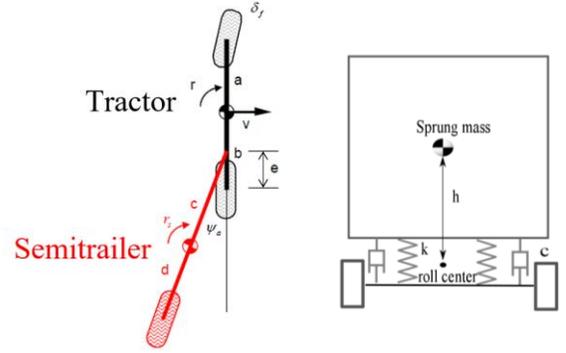

Fig. 4. Lateral-yaw-roll model of articulated truck

The linear model is expressed in the form of (1) for consistency,

$$\begin{aligned} \dot{x} &= f(x) + g(x)\delta_f + g_{d1}(x) r_d + g_{d2}(x) F_y \\ &= Ax + B\delta_f + E_1 r_d + E_2 F_y, \end{aligned} \quad (13)$$

The input to the system is the steering angle $\delta_f$ of the tractor front axle and the disturbances are road curvature $r_d$ and side wind $F_y$, where $r_d$ is the measured disturbance, namely, $d_1$ in (1) and $F_y$ is the unmeasured disturbance, namely, $d_2$ in (1).

A priori, the above linear model is only valid under the following assumptions:

- The longitudinal speed $v_x$ of the truck has small variation;
- Due to the stiff connection on the roll dimension, the roll

- angle of the tractor and semitrailer are the same;
- The pitch and vertical motion are weakly coupled with the lateral, yaw and roll motion, and are ignored in the model;
- The angles are small and therefore the dynamics can be approximated by a linear model.

The simulations performed later in TruckSim support that these assumptions are satisfied in a highway lane keeping scenario.

**Remark 2:** *The methods developed in this paper, including the CBF synthesis, the trajectory optimization, and the continuous-hold controller, all apply to nonlinear models. Hence, for the remainder of the paper, we denote the model as in* (1).

*D. The virtual constraint for the truck model*

We select the lateral displacement with preview as the output for feedback linearization:

$$z = h(x) := y + T_0 v_x \psi \quad (14)$$

with $T_0$ being the preview time, as shown in Fig. 5.

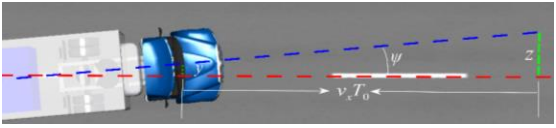

Fig. 5. Preview deviation as output

The output $z$ so-defined has relative degree 2 for any $r_d$, i.e.,

$$\mathcal{L}_g h = 0, \mathcal{L}_g \mathcal{L}_{f+g_{d1}r_d} h \neq 0, \quad (15)$$

To be more specific, the output dynamics is

$$\begin{aligned} z &= h(x), \\ \dot{z} &= \mathcal{L}_f h + \mathcal{L}_{g_{d1}} h \cdot r_d, \\ \ddot{z} &= \mathcal{L}_f^2 h + \mathcal{L}_g \mathcal{L}_f h \cdot u + \mathcal{L}_f \mathcal{L}_{g_{d1}} h \cdot r_d. \end{aligned} \quad (16)$$

By A-1, $r_d$ changes slowly compared to the dynamics, therefore, $\dot{r}_d$ is omitted. Since there are eight states but only $z$ and $\dot{z}$ are used in the feedback linearization, six dimensions of the state space are hidden. It is shown that the zero dynamics of the system is exponentially stable, see Section A in the appendix for detail. Since $\rho = 2$, the feedback structure in (11) is essentially a PD controller:

$$u = -\frac{1}{\mathcal{L}_g \mathcal{L}_f h(x)} \begin{bmatrix} K_p (z - h_{des}) + K_d (\dot{z} - \dot{h}_{des}) + \\ \mathcal{L}_f^2 h(x) + \mathcal{L}_f \mathcal{L}_{g_{d1}} h(x) r_d \end{bmatrix}, \quad (17)$$

where $K_p$ and $K_d$ are the PD gains.

At this point, specifying the desired performance of the truck is simplified to designing $h_{des}$, the desired trajectory of the output $z$, which is discussed in Section IV.

**Remark 3:** *If smooth steering angles are desired, the control design model can be augmented with an integrator appended to $u$. In this case, the system has relative degree three and the control design is nearly the same.*

### III. SYNTHESIS OF CONTROL BARRIER FUNCTION

In this section, we review some existing results for CBFs and present the synthesis process of a CBF for LK control of a truck.

*A. Overview of Control Barrier Function*

Control barrier functions were first proposed in [1] in a reciprocal form and a zeroing CBF was subsequently introduced in [24], which is more robust than the reciprocal form. A zeroing CBF is a scalar function $b(x)$ of the state $x$ that is positive in the safe set, and negative in the danger set. The algebraic set $\{x \mid b(x) = 0\}$ is called the boundary of the CBF. For a zeroing CBF, the barrier condition can be written as

$$\dot{b} + \kappa \alpha(b) \geq 0, \quad (18)$$

where $\kappa > 0$ is a positive constant, and $\alpha$ is an extended class $\mathcal{K}$ function, that is, a function $f: \mathbb{R} \to \mathbb{R}$ satisfying

- $f$ is strictly increasing;
- $f(0) = 0$.

When $b(x) > 0$, $\dot{b}$ can be negative, but is lower bounded by $-\kappa\alpha(b)$; at the boundary, $\dot{b}$ should be nonnegative, which makes the set $\{x \mid b(x) \geq 0\}$ controlled invariant. When $b(x) < 0$, the condition in (18) enforces convergence to the set $\{x \mid b(x) \geq 0\}$ by setting a lower bound $\dot{b} \geq -\kappa\alpha(b) > 0$.

*B. Synthesis of CBF using Sum of Squares programming*

The synthesis of a CBF is nontrivial. We use the Sum of Squares (SOS) technique to synthesize a CBF for the truck LK problem.

SOS has been widely used in the computation of invariant sets and barrier certificates for continuous dynamic systems, and it can be efficiently solved with semidefinite programming (SDP). In addition, with the help of Putinar's PositivStallensatz, SOS condition is enforced on semialgebraic sets via multipliers [25]. For more information, see [2, 26-31].

We focus on a dynamic system with the control affine structure in (1), where the dynamics assumed to be polynomial and $\mathcal{U}, \mathcal{D}$ are known semialgebraic sets:

$$\mathcal{U} = \{u \mid h_u(u) \geq 0\}, \mathcal{D} = \{d \mid h_d(d) \geq 0\}. \quad (19)$$

To make the notation compact, let $g_d(x) = [g_{d1}(x), g_{d2}(x)]$, $d = [d_1, d_2]^T$. In CBF synthesis, we set $\alpha(b) = b$, and seek a polynomial CBF $b(x)$ that satisfies the following:

$$\{x \mid b(x) \geq 0\} \cap X_d = \emptyset; \quad (20)$$

$$\begin{aligned} &\forall x \in \{x \mid b(x) \geq 0\}, \forall d \in \mathcal{D}, \exists u \in \mathcal{U}, s.t. \\ &\frac{db}{dx}(f(x) + g(x)u + g_d(x)d) + \kappa b \geq 0, \end{aligned} \quad (21)$$

where $X_d$ is the danger set, a semialgebraic set of $x$:

$$X_d = \{x \mid h_{xd}(x) \geq 0\}, \quad (22)$$

$\kappa > 0$ is a positive constant, and condition (21) is referred to as

the **CBF condition**.

The difference between a barrier certificate and a CBF shows up in condition (21), which depends on the control input, $u$. The existential quantifier of $u$ renders (21) not directly solvable by current SOS solvers and thus we seek a conservative approximation, in which we assume the control input $u$ comes from a polynomial controller of $x$ and $d$, namely,

$$\begin{aligned}
&\{x|\ b(x) \geq 0\} \cap X_d = \varnothing; \\
&\forall x \in \{x|\ b(x) \geq 0\}, u(x,d) \in \mathcal{U}; \\
&\forall x \in \{x|\ b(x) \geq 0\}, \forall d \in \mathcal{D}, \\
&\frac{db}{dx}\big(f(x) + g(x)u(x,d) + g_d(x)d\big) + \kappa b(x) \geq 0.
\end{aligned} \quad (23)$$

The input may depend on measured disturbance, but not on unmeasured disturbance.

Even with the simplification, there are two bilinear terms that must be addressed to make the problem solvable by SOS. The first bilinear term is between $b(x)$ and $u(x,d)$. We use bilinear alternation [2, 32], which iterates the following two steps [2]:

- Fix the barrier candidate, search for a controller;
- Fix the controller, search for a better barrier candidate.

The following SOS program solves for a controller with a fixed $b(x)$:

min $e$ s.t.
$$\begin{aligned}
&h_u\big(u(x,d)\big) - s_1(x,d)b(x) - \sum_i s_2^i(x,d)h_d^i(d) \in \Sigma[x,d] \\
&\frac{db}{dx}\big(f(x) + g(x)u(x,d) + g_d(x)d\big) + \kappa b(x) + s_3(x,d)b(x) \quad (24)\\
&- \sum_i s_4^i(x,d)h_d^i(d) + eQ(x,d) \in \Sigma[x,d] \\
&s_1, s_2, s_3, s_4 \in \Sigma[x,d],
\end{aligned}$$

where $s_1$, $s_2$, $s_3$, $s_4$ are the SOS multipliers. $Q$ is a fixed SOS polynomial of $x$ and $d$; $e$ is a relaxation scalar variable that makes this SOS program feasible. When $e \leq 0$, (24) is a sufficient condition of (23). $s_3$ is used to enforce the CBF condition only when $b(x) \geq 0$. The first SOS constraint restricts the input to be bounded by $\mathcal{U}$; the second SOS constraint enforces the CBF condition.

**Remark 4:** *The choice of $Q$ depends on the order of the polynomial required to be SOS. In many cases, $Q$ can simply be $x^T x$.*

The other step of the bilinear alternation searches for a better CBF candidate with the controller held fixed. In this step, the second bilinear term emerges. Because the CBF condition is enforced only when $b(x) \geq 0$, an SOS multiplier is used to enforce this condition, which creates a bilinear term between $b$ and the multiplier. We use perturbation to solve this bilinear term. The idea is to enforce the CBF condition inside the 0-level set of the current CBF candidate $b_0$, and search for a small perturbation $\Delta b$, as shown in (25).

min $e$ s.t.
$$\begin{aligned}
&-b_0(x) - \Delta b(x) - \sum_i s_1^i(x)h_{xd}^i(x) \in \Sigma[x] \\
&\frac{d(b+\Delta b)}{dx}\big(f(x) + g(x)u(x,d) + g_d(x)d\big) + \kappa(b_0 + \Delta b)(x) \quad (25)\\
&- \sum_i s_2^i(x,d)h_d^i(x) + s_3(x,d)b_0(x) + eQ(x,d) \in \Sigma[x,d] \\
&s_1 \in \Sigma[x],\ s_2, s_3 \in \Sigma[x,d],\ \|\Delta b\| \leq \epsilon \|b_0\|.
\end{aligned}$$

The norm is taken on the coefficient of $\Delta b$ and $b_0$ for some selected monomial bases. Note that the CBF condition is enforced on the zero level set of $b_0$ rather than $b$ which makes the bilinear term disappear (since $b_0$ is fixed and not part of the SOS variables). Because of this, we need the zero level set of $b_0 + \Delta b$ to be similar to that of $b_0$, which is enforced by the last constraint, with $1 \gg \epsilon > 0$, a constant that keeps $\Delta b$ small compared to $b_0$. The algorithm iteratively updates $b_0$ by $b_0 + \Delta b$ until no further progress can be made. Upon convergence, that is, $\Delta b \to 0$, the original CBF condition is enforced.

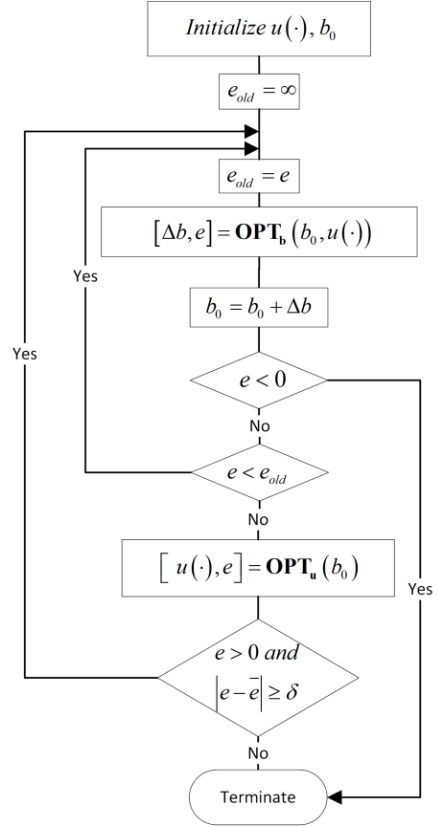

Fig. 6. Synthesis of a CBF via SOS

In summary, there are two loops in the algorithm. The inner loop iterates the perturbation process, updating $b_0$ with $b_0 + \Delta b$ while the outer loop iterates between updating $b$ and updating $u(\cdot)$. Denote the optimization in (24) as $[u(\cdot), e] = \mathbf{OPT_u}(b)$, with $b$ as input, $u(\cdot)$ and $e$ as output; and denote the optimization in (25) as $[\Delta b, e] = \mathbf{OPT_b}(b, u(\cdot))$, with $b(x)$ and $u(\cdot)$ as input, $\Delta b$ and $e$ as output. The iteration terminates when a valid CBF is found or no improvement can be made, as shown in Fig. 6. Some key parameters for the CBF of lane keeping are listed in *TABLE I*.

TABLE I  LIST OF PARAMETERS

| | |
|---|---|
| $v_x$ | $20m/s$ |
| Bound on $y$ | $\pm 0.3m$ |
| Bound on $\phi$ | $\pm 0.1 rad$ |
| Bound on $r_d$ | $\pm 0.02 rad/s$ (turning radius of 1000m) |
| Bound on $F_y$ | $\pm 2000N$ |
| Bound on $\delta_f$ | $\pm 0.2 rad$ |

## IV. TRAJECTORY OPTIMIZATION

Although a CBF guarantees safety of the system's trajectories, the closed-loop performance could be compromised if the student controller is not properly designed. For example, in Fig. 15 we show a student controller designed with LQR requiring frequent interventions from the CBF and thus leading to bad ride comfort. In this chapter, we present an optimization procedure that incorporates the CBF condition, which is then used to train a student controller that is compatible with the CBF. In addition to the CBF condition, other constraints are needed to ensure the stability of the continuous hold controller, as introduced later in Section V.A.

### A. Direct Collocation

As discussed in Section II, the trajectory optimization is boiled down to the optimization of $h_{des}$, the desired trajectory of the output $z$. Direct collocation is used to generate the trajectory of the states and $h_{des}$, while $h_{des}$ is imposed as the virtual constraint.

Direct collocation is widely employed in trajectory optimization problems due to its effectiveness and robustness and is capable of enforcing nonlinear and nonconvex constraints. It is thus chosen to optimize the trajectory while enforcing the virtual constraint. It works by replacing the explicit forward integration of the dynamical systems with a series of defect constraints via implicit Runge-Kutta methods, which provides better convergence and stability properties particularly for highly underactuated dynamical systems. The result is a nonlinear programming problem (NLP) [33].

In this paper, we utilize a modified Hermite-Simpson scheme based direct collocation trajectory optimization method [34]. Particularly, the flow (a.k.a. trajectory), $x(t)$, of the continuous dynamical system in (13) is approximated by discrete value $x^i$ at uniformly distributed discrete time instant $0 = t_0 < t_1 < t_2 < \cdots < t_N = T$ with $N > 0$ being the number of discrete intervals. Let $x^i$ and $\dot{x}^i$ be the approximated states and first order derivatives at node $i$, they must satisfy the system dynamic equation given in (13). Further, if these discrete states satisfy the following defect constraints at all interior points $i \in [1,3,\ldots,N-1]$,

$$\zeta^i := \dot{x}^i - \frac{3N}{2T}\left(x^{i+1} - x^{i-1}\right) + \frac{1}{4}\left(\dot{x}^{i-1} + \dot{x}^{i+1}\right) = 0,$$
$$\delta^i := x^i - \frac{1}{2}\left(x^{i+1} + x^{i-1}\right) - \frac{T}{8N}\left(\dot{x}^{i-1} - \dot{x}^{i+1}\right) = 0, \quad (26)$$

then they are accurate approximations of the given continuous dynamics. (26) defines the modified Hermite-Simpson conditions for the direct collocation trajectory optimization [34].

Based on the above formulation, now we can construct a constrained nonlinear programming problem to solve the trajectory optimization with the virtual constraint for the articulated truck model. To incorporate the virtual constraints based feedback control with the trajectory optimization, we enforce the output dynamics equation given in (16) at each node. Then the control input $u^i$ will be implicitly determined via this constraint without explicitly enforcing it as in (17). Further, the output $z$ and its derivative $\dot{z}$ should equal to the desired trajectory $h_{des}(t)$ at $t=0$ to ensure that the system lies on the zero dynamics manifold $\forall t \in [0,T]$.

The desired trajectory $h_{des}$ is parameterized as Bezier curve, which is widely used in computer graphics and related fields. A Bezier curve of order $m$ is an $m$-th order polynomial defined on $[0,1]$:

$$\mathbf{B}(s) = \sum_{i=0}^{m} \alpha_i \binom{m}{i} s^i (1-s)^{m-i}, \quad (27)$$

where $\alpha_i$ are the Bezier coefficients. [d] The Bezier order is chosen to be 8.

Let $\mathcal{J}(\cdot)$ be the cost function to be minimized, the trajectory optimization problem can be stated as:

$$\begin{aligned}
&\arg\min \ \mathcal{J}(\cdot) \quad s.t. \\
&\zeta^i = 0, \delta^i = 0, \\
&\dot{x}^i = f(x^i)x^i + g(x^i)u^i + g_{d1}(x^i)d_1, \\
&\ddot{z}^i - \ddot{h}_{des}(t_i) + K_p\left(z^i - h_{des}(t_i)\right) + K_d\left(\dot{z}^i - \dot{h}_{des}(t_i)\right) = 0, \\
&x(t_0) = x^0, \\
&z^0 - h_{des}(t_0) = 0, \\
&\dot{z}^0 - \dot{h}_{des}(t_0) = 0, \\
&-u_{\max} \le u^i \le u_{\max}, \\
&\dot{b}\left(x^i,\dot{x}^i\right) + \kappa \frac{e^{b(x^i)}-1}{e^{b(x^i)}+1} \ge 0, \\
&V\left(x(T) - \eta_x(d_1)\right) \le c_1 V\left(x^0 - \eta_x(d_1)\right), \\
&\|x_2(T) - \gamma(x)(T)\| \le c_3,
\end{aligned} \quad (28)$$

where $z^i = z(x^i)$, $\dot{z}^i = \dot{z}(x^i)$, and $\ddot{z}^i = \ddot{z}(x^i,\dot{x}^i)$, respectively. The first 3 lines of constraints correspond to the colocation constraint; 4th line specifies the initial states; 5th and 6th line correspond to the virtual constraint; the 7th line is the input constraint; the 8th line is the CBF constraint, the last two constraints are needed to guarantee stability of the continuous hold controller, which will be explained in Appendix Section B.

---

[d] Bezier curve can parameterize trajectories of any finite length by scaling the input. Suppose the horizon of $h_{des}$ is $T$, then the input is defined as $s = t/T$.

**Remark 5**: *The CBF condition is modified based on (18). Since $\frac{e^b - 1}{e^b + 1}$ is bounded within $[-1, 1]$, when $b(x)$ is small, the lower bound for $\dot{b}$ saturates at 1, instead of growing linearly as $-\gamma b$, which may be too difficult to satisfy. Besides, when $b(x) = 0$, $\frac{e^b - 1}{e^b + 1} = 0$, which resembles the original CBF condition in (20). Since $\frac{e^b - 1}{e^b + 1}$ is still an extended class $\mathcal{K}$ function, by Proposition 1 in [35], $\{x \mid b(x) \geq 0\}$ is still invariant under the modified constraint.*

The cost function in (28) is a weighted sum of multiple cost functions, consisting of the following terms:

- Final value cost $V(x^T - \eta_x(r_d))$, where $\eta_x(r_d)$ is the steady state under a given $r_d$, and $V(\cdot)$ is a Lyapunov function around the origin.
- $\int z^2 dt$, the square integral of $z$
- $\int \dddot{z}^2 dt$, the square integral of jerk
- $\|y\|_\infty$, the maximum deviation from road center
- $\|r\|_\infty$, maximum yaw rate
- $\int u^2 dt$, the square integral of the input
- $|\alpha_m|$, penalty on the last Bezier coefficient (facilitate convergence of the Bezier curve)

The terms that consist of function integrals are approximately computed using the Simpson's quadrature rule [36].

The setting of the constraints and costs seem complicated, they are the result of repeated trial and tuning. It should be emphasized that CBF constraint is enforced in the trajectory optimization. We hope that by enforcing CBF condition on the training set, the policy generated by supervised learning inherits this property.

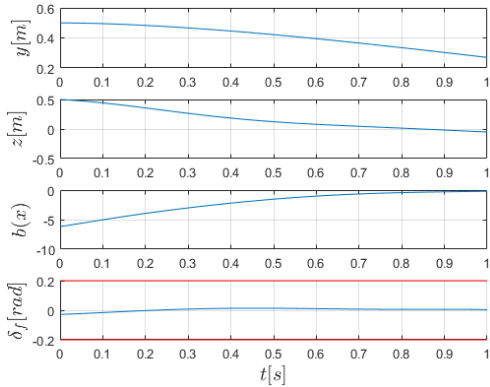

Fig. 7. Example of trajectory optimization result

Fig. 7 shows an example trajectory with initial lateral deviation $y_0 = 0.5 m$ and road yaw rate $r_d = 0.02 rad/s$. The plot of $y$ and the Bezier output $z$ shows that the trajectory is converging to the lane center. The plot of the CBF value and the control input shows that the trajectory generated by direct collocation satisfies the input and CBF constraints.

The trajectory optimization is solved with FROST, which uses a symbolic calculation to boost the nonlinear optimization [37]. The trajectory optimization for each initial condition can be finished within 10 seconds.

### B. Generating the training set

It is impossible to perform trajectory optimization for all the initial conditions offline, so instead, we use supervised learning to train the mapping from initial conditions to desired trajectories with a finite trajectory library, which is generated by the above-described trajectory optimization process.

By varying the initial conditions and generating the corresponding trajectories with direct collocation, we hope to 'train' the neural network to generate good trajectories for various initial conditions. The inputs to the neural network are called features, denoted as $\Phi$; in our case, they are variables that describe the initial condition. The output of the neural network is a vector of control parameters, denoted as $\mathcal{A}$, in this case, the Bezier coefficients.

$$\pi : \Phi \to \mathcal{A}. \qquad (29)$$

The selection of initial conditions is done in a grid fashion. We define a grid on the feature space and perform trajectory optimization on each of the grid points. Since the zero dynamics is stable, $z(t) \to h_{des}(t)$ for $h_{des} \in C^2$ implies $x(t) \to x_{des}(t)$, where $x_{des}$ is the desired state trajectory corresponding to $h_{des}$. This implies that we only need two states to determine the asymptotic behavior of the system, but not necessarily the transient behavior. In practice, the more states we use to parameterize the initial condition, the finer the trajectory library will be.

However, under a grid fashion of drawing samples, the number of samples needed grows exponentially with the state dimension. Therefore, the dimension of $\Phi$ is limited by available computation power. We let $\Phi$ contain 6 features, including 5 states and $r_d$:

$$\Phi = [y, \psi, r, \psi_a, v_y, r_d]. \qquad (30)$$

Under this setup, the computation needed to generate the trajectory library is manageable (about 20 hours on a desktop). With more computation power, a higher dimensional $\Phi$ can lead to a finer trajectory library.

Even though most driving behavior is mild, it is important that the controller be able to handle bad initial conditions. We generate, therefore, two training sets, denoted as $S_1$ and $S_2$, where $S_1$ consists of trajectories defined for a duration of 1 second, and the features of the trajectories have a wider span, and $S_2$ consists of trajectories defined over a 3 second window, with the features more concentrated around the origin. $S_1$ is used to train a mapping for severe initial conditions and transients, and $S_2$ is used to train a mapping for mild situations and normal driving. Some of the initial conditions might render the trajectory optimization infeasible, therefore only the feasible cases are included in the training sets. In the implementation, the CH controller will choose which mapping to use based on the severity of the situation.

TABLE II  TRAINING SET PARAMETER SETTING

| Feature | $S_1$ | $S_2$ |
|---|---|---|
| $y$ range | $[-0.5, 0.5][m]$ | $[-0.3, 0.3][m]$ |
| $v_y$ range | $[-1, 1][m/s]$ | $[-1, 1][m/s]$ |
| $\psi$ range | $[-0.04, 0.04][rad]$ | $[-0.04, 0.04][rad]$ |
| $r$ range | $[-0.06, 0.06][rad/s]$ | $[-0.03, 0.03][rad/s]$ |
| $r_d$ range | $[-0.03, 0.03][rad/s]$ | $[-0.025, 0.025][rad/s]$ |
| $\psi_a$ range | $[-0.04, 0.04][rad]$ | $[-0.04, 0.04][rad]$ |

The parameters for the training are included in *TABLE II*. In total, there are 62825 trajectories in $S_1$, and 29300 trajectories in $S_2$.

*C. Supervised learning*

With the training set ready, there are several choices for the supervised learning, such as linear regression, Gaussian process regression, and neural networks. In our problem, since there is no structural information about the trajectory generator and we need strong expressive power to capture the potentially complicated mapping from the initial condition to the desired trajectory, we choose a neural network for its strong expressive power.

We train a neural network that has 6 hidden layers with 200 neurons in each layer and use the ReLU function as the rectifier. The training is performed using Tensorflow [38]. 85% of the data is used for training and 15% is used for testing. TABLE III shows the mean squared error (MSE) of the training result.

TABLE III  TRAINING RESULT

|  | $S_1$ | $S_2$ |
|---|---|---|
| MSE of training data | 0.13 | 0.0023 |
| MSE of testing data | 0.16 | 0.0024 |

V. IMPLEMENTATION OF LEARNING BASED CONTROLLER

*A. Continuous hold feedback control*

Once the trajectory generator is trained, we can generate a finite horizon desired trajectory for a given initial condition. In order to piece together the finite horizon trajectories and synthesize a controller from the trajectory generator, we employ a continuous hold (CH) controller. The name continuous hold comes from the analogy with a zero-order hold and an *n-th* order hold. While an *n-th* order hold approximates the segment between two consecutive sampling times with an *n-th* order polynomial, continuous hold executes a predefined continuous trajectory.

The idea of continuous hold is not claimed to be novel; a motion primitive is a special type of continuous hold [39]. The trajectory is updated in an event-triggered fashion, which will be discussed in detail in Section V.B. While event-triggered finite-horizon control is studied in [40], in the CH setting, it should be noted that the control action between triggering events is a continuous function of time and states instead of being a constant.

For the truck example, the basic continuous hold controller [41] must be extended to systems with exogenous disturbances. The stability and set invariance property of the CH controller are proved, including the analysis for the case when only a subset of the state is used for feedback, in Appendix Section B.

*B. Event-triggered update of the CH controller*

The CH controller uses the mapping trained by supervised learning to generate a desired trajectory $h_{des}$ for the output $z$ based on the current state and $r_d$, then track the desired trajectory with the control law in (17). The desired trajectory will be updated under three circumstances:
- The desired trajectory is executed to the end
- There is a significant change in road curvature
- The trajectory tracking error becomes large

In the first case, since the trajectory optimization has a finite horizon (1s or 3s), the neural network will use the current value of the features to generate a new desired trajectory. In the second case, if the road curvature $r_d$ differs much from that used to generate the current desired trajectory, the trajectory should be updated since $r_d$ is assumed to be constant during the entire horizon of the trajectory. The rest of the features are simply initial conditions, so their change does not trigger an update of the desired trajectory. In the third case, when the trajectory deviates too far from the desired trajectory, re-planning is called for. This is likely to be caused by an unexpected disturbance, such as wind gust.

When switching from one trajectory to the next, smoothing is performed to make sure that $h_{des}$ is twice differentiable, which ensures that the control signal is continuous. The smoothing process is explained in the Appendix Section C.

*C. CBF as a supervisory controller*

Even though the CBF condition is enforced in the trajectory optimization used in the training set, after supervised learning, there is no guarantee that the trajectory generated by the neural network always satisfies the CBF condition. Therefore, CBF is still implemented as a supervisory controller on top of the CH controller, as shown in Fig. 2. The CBF solves the following optimization:

$$\min_{\Delta u} w_1 \|\Delta u\|^2 + w_2 \|\Delta u - \Delta u_{old}\|^2 \\ s.t. \mathcal{C}(x, d, u_0 + \Delta u) \geq 0, u_0 + \Delta u \in \mathcal{U}, \quad (31)$$

where $\Delta u$ is the intervention of the CBF, $\Delta u_{old}$ is the intervention of the previous time instant, $\mathcal{C}(\cdot)$ is the CBF condition. The reason for the second penalty term is to prevent chattering if intervention is necessary. The CBF condition is defined as

$$\begin{cases} \dot{b} + \gamma b \geq 0, & \text{if } b(x) \geq 0 \\ \dot{b} + \gamma \dfrac{e^b - 1}{e^b + 1} \geq 0, & \text{if } b(x) \leq 0 \end{cases}, \quad (32)$$

where the transition at $b(x)=0$ is continuous, i.e. the two constraint coincides at $b(x)=0$.

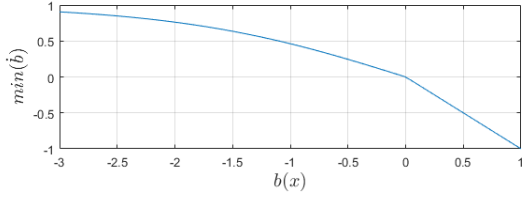

Fig. 8. lower bound for $\dot{b}$

**Remark 6**: *When $b(x) \geq 0$, the existence of $\Delta u$ is guaranteed by the construction of the CBF; when $b(x) \leq 0$, there is no guarantee of feasibility. When (31) is infeasible, the input is saturated by $\mathcal{U}$.*

## VI. SIMULATION RESULT

We validate our control design on TruckSim, a high fidelity physics-based simulation software that is widely acknowledged by the trucking industry. The model picked for simulation has 312 states and is a tractor-semitrailer with heavy cargo in the trailer, weighing 35 tons in total; see Fig. 9.

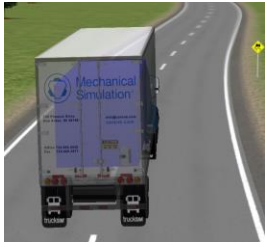

Fig. 9. Animation with a 312 state model in TruckSim

The truck is asked to drive on a road with a minimum turning radius of 1000 m at 20m/s. A side-wind is simulated as a lateral force and roll moment to the truck. Because of the heavy cargo, the truck has a high CG. Hence, the roll motion in the simulation is significant and the commanded maneuvers are aggressive.

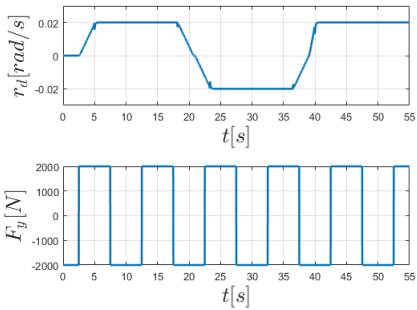

Fig. 10. Disturbance to the system

As shown in Fig. 10, the road profile consists of segments with constant curvature (per US road design standards) Though rather extreme, the side-wind is a square wave with maximum allowed magnitude.

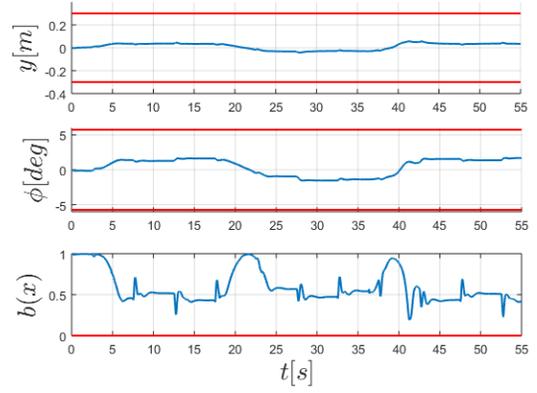

Fig. 11. Value of the CBF and key states during simulation

Fig. 11 shows the value of the CBF and two key states. Lateral deviation $y$ and roll angle $\phi$ never exceed the desired limits (plotted in red) and $b(x)$ was always above zero, showing that the CBF (safety) bound was never breached.

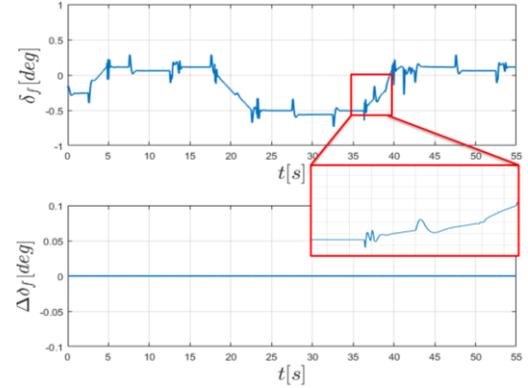

Fig. 12. Input and intervention of CBF during simulation

The steering input trajectory is shown in Fig. 12. We zoom in the input to show a 5 second period of input. The input is actually reasonably smooth. The bumps are necessary to counter the side-wind when it changes direction. The lower plot shows $\Delta \delta_f$, and its constant value of zero indicates that no interventions from CBF occurred.

To demonstrate the controller's ability to handle bad initial conditions, we perturb the lateral deviation with a square wave, simulating the situation when the initial position is 0.5m from the lane center, as shown in Fig. 13 and Fig. 14.

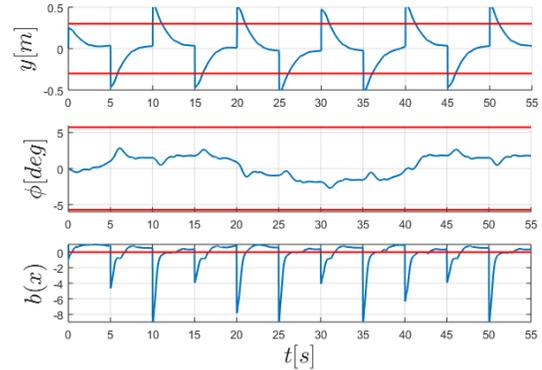

Fig. 13. Value of CBF and key states with large initial deviations

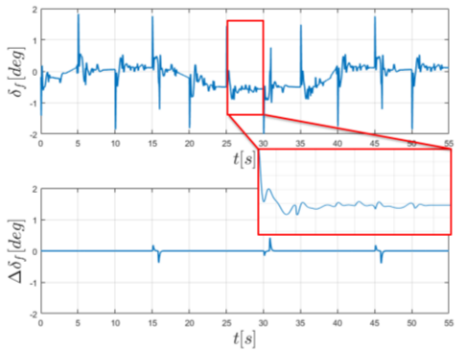

Fig. 14. Input and intervention of CBF with large initial deviations

Fig. 14 shows the input under a large deviation. The CBF intervened 3 times, and the interventions are mild compared to the size of $u_0$. When $b(x)$ was below zero, the learned controller was able to drive the system back to the safe set without the intervention of the CBF.

As a comparison, we tuned an LQR controller with feedforward control of $r_d$, and it performed very well under normal driving conditions. However, when the initial condition is bad (under the same setting as Fig. 13), the LQR controller triggered intervention from the CBF multiple times (11 times) and the jerk was severe, as shown in Fig. 15.

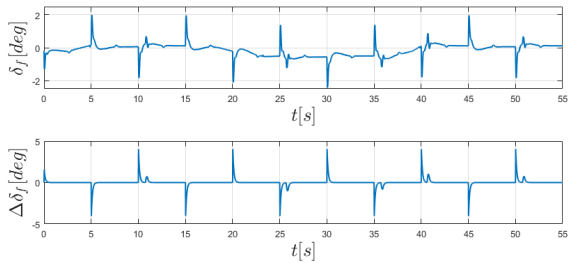

Fig. 15. Simulation result with LQR as student controller

Though the LQR controller was fine-tuned, it triggered severe intervention from the CBF frequently. On the other hand, we observed none or very mild interventions from the CBF under the learning-based controller in all trial simulations when the states are within the span of the training set.

## VII. CONCLUSION AND DISCUSSION

We propose a supervised learning approach to construct controllers with smooth performance and a provable safety guarantee. The idea is to use trajectory optimization to generate a training set consisting of trajectories that satisfy a Control Barrier Function (CBF) safety constraint, then use supervised learning to extract a mapping from system initial conditions to desired trajectories. The policy generated with supervised learning inherits the good properties of the training set, though nothing can be proved. On top of that, a safety guarantee is formally imposed with a CBF as a supervisory controller. The simulations showed that the proposed approach is able to reduce the intervention of the CBF and therefore provide high-quality closed-loop performance while guaranteeing safety.

We chose to learn a mapping from initial conditions to the desired output trajectory, instead of a mapping from the initial condition to the desired input trajectory. Trajectory tracking was implemented with a continuous-hold (CH) controller. The CH control structure is able to transform the synthesis problem into a trajectory optimization problem, which may be much simpler for complicated nonlinear systems such as trucks and robots [41].

There are problems to be solved for the proposed method. First, when the initial condition is not contained inside the feature range of the training set, i.e. when the neural network is doing extrapolation rather than interpolation, the performance can be poor. Though rather obvious, it is important to emphasize that to obtain good performance over a wide range, one needs to have training data with adequate coverage. Second, when training data from a large range of features are stacked together, the regression accuracy may drop and the performance suffers. To solve this, one might need more a complicated neural network structure, or use multiple neural networks for different situations.

## ACKNOWLEDGMENT

The work of Yuxiao Chen, A. Hereid, and Huei Peng is supported by NSF Grant CNS-1239037. The work of J. Grizzle is supported by Toyota Research Institute (TRI).

## APPENDIX

### A. Zero dynamics of the truck lateral dynamics

To show the zero dynamics, we use the following state transformation that renders a new choice of states:

$$\chi = T = \begin{bmatrix} z & \dot{z} & \sigma^T \end{bmatrix}^T, \quad (33)$$

where $T$ is a full-rank linear transformation matrix:

$$T = \begin{bmatrix} 1 & 0 & v_x T_0 & 0 & 0 & 0 & 0 & 0 \\ 0 & 1 & v_x & v_x T_0 & 0 & 0 & 0 & 0 \\ 0 & 0 & 1 & 0 & 0 & 0 & 0 & 0 \\ 0 & -B^4 & 0 & B^2 & 0 & 0 & 0 & 0 \\ 0 & 0 & 0 & 0 & 1 & 0 & 0 & 0 \\ 0 & -B^6 & 0 & 0 & 0 & B^2 & 0 & 0 \\ 0 & 0 & 0 & 0 & 0 & 0 & 1 & 0 \\ 0 & -B^8 & 0 & 0 & 0 & 0 & 0 & B^2 \end{bmatrix}, \quad (34)$$

and $B^i$ is the $i$-th entry of $B$ in (13). The transformation is linear and full rank. The dynamics under $\chi$ is

$$\dot{\chi} = \overline{A}\chi + \overline{B}u + \overline{E}r_d, \quad (35)$$

Moreover,

$$\overline{B} = TB = \begin{bmatrix} 0 & CAB & \mathbf{0}_{1\times 6} \end{bmatrix}^T, \quad (36)$$

where $C = \begin{bmatrix} 1 & 0 & T_0 v_x & \mathbf{0}_{1\times 5} \end{bmatrix}$. Therefore, the dynamics under $\chi$ can be written as

$$\dot{\chi} = \begin{bmatrix} \dot{\sigma} \\ \dot{z} \\ \ddot{z} \end{bmatrix} = \begin{bmatrix} \Gamma(\sigma, z, \dot{z}) \\ \dot{z} \\ CA^2 x \end{bmatrix} + \begin{bmatrix} \mathbf{0}_{6\times 1} \\ 0 \\ CAB \end{bmatrix} u + \begin{bmatrix} \mathbf{0}_{6\times 1} \\ 0 \\ CAE_1 \end{bmatrix} r_d, \quad (37)$$

where $\Gamma(\sigma, z, \dot{z})$ has exponentially stable zero dynamics, meaning $\dot{\sigma} = \Gamma(\sigma, 0, 0)$ is exponentially stable.

*B. Analysis of the continuous hold controller*

In this section, we present the stability and set invariance analysis of the continuous hold (CH) controller.

*B.1 CH controller with full state parameterization*

Systems of the following form are considered:
$$\dot{x} = f(x, u, d), \qquad (38)$$

where $x$, $u$ and $d$ are the state, the input and the measured disturbance respectively.

**Remark 7**: *The result about the CH controller is applicable to general nonlinear dynamic model; the control-affine nonlinear model in (1) (assuming $d_2 = 0$) and the linear model of the truck are special cases.*

We make the following assumptions about the system dynamics.

**A-5**: $f : \mathbb{R}^n \times \mathbb{R}^m \times \mathbb{R}^p \to \mathbb{R}^n$ *is locally Lipschitz continuous in $x$, $u$ and $d$.*

**A-6**: $\exists \mathcal{D} \subseteq \mathbb{R}^p$ *and mappings* $\eta_x : \mathcal{D} \to \mathbb{R}^n$, $\eta_u : \mathcal{D} \to \mathbb{R}^p$, *Lipschitz continuous, such that* $f(\eta_x(d), \eta_u(d), d) = 0$, *i.e., for every exogenous disturbance $d \in \mathcal{D}$, there exist two mappings $\eta_x$ and $\eta_u$ that map any $d \in \mathcal{D}$ to a unique equilibrium point and a unique input that maintains the equilibrium.*

**Remark 8**: *There may be non-unique equilibrium points of the system due to the cyclic coordinates, i.e., states that do not affect the dynamics, see [42] for detail. Therefore, a function $\eta_x$ is needed to select a single equilibrium point given $d$. A-3 gives a possible definition of $\eta_x$.*

**A-7**: $\forall d \in \mathcal{D}$, *there is an open ball $B_d \subset \mathbb{R}^n$ about the origin, and a positive-definite, locally Lipschitz-continuous function $V_d : B_d \to \mathbb{R}$, and constants $0 \leq \alpha_1 \leq \alpha_2$ such that* $\forall x \in B_d + \eta_x(d)$,
$$\begin{aligned} \bar{x} &= x - \eta_x(d), \\ \alpha_1 \bar{x}^T \bar{x} &\leq V_d(\bar{x}) \leq \alpha_2 \bar{x}^T \bar{x}. \end{aligned} \qquad (39)$$

**A-8**: $\exists \mathcal{S} \subseteq \mathbb{R}^n$, *compact, such that* $\forall d \in \mathcal{D}, \eta_x(d) \in \mathcal{S}$. *There exists CBF $b(x)$, such that* $\forall x \notin \mathcal{S}, b(x) \leq 0$; $\forall d \in \mathcal{D}$, $b(\eta_x(d)) > 0$. *Moreover,* $\forall \xi \in \mathcal{S}, \forall d \in \mathcal{D}$, *there exists $u^d_\xi : [0, T_p] \to \mathbb{R}^m$ and a corresponding state trajectory $\varphi^d_\xi : [0, T_p] \to \mathbb{R}^n$ satisfying*
$$V_d\left(\varphi^d_\xi(T_p) - \eta_x(d)\right) \leq c_1 V_d\left(\varphi^d_\xi(0) - \eta_x(d)\right)$$
$$\left.\frac{db}{dx}\right|_{\varphi^d_\xi(t)} \dot{\varphi}^d_\xi(t) + \kappa b\left(\varphi^d_\xi(t)\right) \geq 0 \qquad (40)$$
$$\forall t \in [0, T_p], \lim_{\xi \to \eta_x(d)} \varphi^d_\xi(t) = \eta_x(d), \lim_{\xi \to \eta_x(d)} u^d_\xi(t) \equiv \eta_u(d),$$

where $T_p > 0$ is the horizon, $\kappa > 0$, $1 > c_1 > 0$ are predefined constants.

A CH controller maintains a timer $\hat{t}$ that is reset to 0 when the triggering event occurs and the desired trajectory is updated. In between events, the timer increases at a constant rate equal to 1. An update is triggered when either the trajectory is executed to its end, i.e. $\hat{t} = T_p$, or when an interruption is detected. A possible interruption includes a change in $d$ or an unexpected disturbance that makes the tracking error too big. The CH input is
$$u^{CH}(\hat{t}, x, d) = u^d_\xi(\hat{t}) + u^{fb}\left(x, \varphi^d_\xi(\hat{t})\right), \qquad (41)$$

where $\xi$ is the initial state when $\hat{t} = 0$, and $u^{fb} : \mathbb{R}^n \times \mathbb{R}^n \to \mathbb{R}^m$ is a feedback controller that tracks $\varphi^d_\xi$.

The closed-loop system under CH feedback is then
$$\dot{x} = f^{CH}(\hat{t}, x, d) := f\left(x, u^d_\xi(\hat{t}) + u^{fb}\left(x, \varphi^d_\xi(\hat{t})\right), d\right). \qquad (42)$$

**A-9**: *For any trajectory $\varphi^d_\xi$ in A-8 that a CH controller tries to follow, there exists a feedback controller $u^{fb} : \mathbb{R}^n \times \mathbb{R}^n \to \mathbb{R}^m$ that makes $\varphi^d_\xi$ uniformly locally exponentially stable, i.e., the closed-loop system in (42) satisfies*
$$\exists B \subseteq \mathbb{R}^n, s.t. \forall 0 \leq t_1 \leq t_2 \leq T_p, \left(x(t_1) - \varphi^d_\xi(t_1)\right) \in B,$$
$$\left\|x(t_2) - \varphi^d_\xi(t_2)\right\| \leq e^{-c_2(t_2 - t_1)} \left\|x(t_1) - \varphi^d_\xi(t_1)\right\| \qquad (43)$$

*for some $c_2 > 0$.*

Next, we present the result on the stability property of the CH controller. First, consider the case when $d$ is fixed.

**Theorem 1**: *Under assumptions A-5, A-6, A-7, A-8, and A-9, for an initial condition $\xi \in \{x \mid b(x) \geq 0\}$ the closed-loop system in (42) will stay inside $\{x \mid b(x) \geq 0\}$, and if $d$ stops changing after $T \geq 0$, the state will converge to $\eta_x(d)$ exponentially.*

*Proof:* From A-8, since $\xi \in \{x \mid b(x) \geq 0\}$, $\xi \in \mathcal{S}$, which means the feedback control in (41) is well defined. From (40), the CBF $b(x)$ remains nonnegative as discussed in Section III, which proves that the state will stay inside $\{x \mid b(x) \geq 0\}$, and thus $x(t) \in \mathcal{S}, \forall t \geq 0$.

When $d$ stops changing, from the Lyapunov condition in A-8, $V_d\left(x(nT_p) - \eta_x(d)\right) \leq c_1^{n-1} V_d\left(x(0) - \eta_x(d)\right)$, $n = 1, 2, 3, \ldots$, which implies $\lim_{n \to \infty} V_d\left(x(nT_p)\right) = 0$. From A-7, the sequence $x(nT_p)$ converges to $\eta_x(d)$. Therefore, from the last assumption in (40), the state stays at $\eta_x(d)$.

∎

## B.2 State decomposition and dimension reduction

As discussed in Section IV.B, under a grid fashion sampling of the initial condition, the computation power limits the dimension of the feature that describes the initial condition. To parameterize the initial condition with a subset of states, we decompose the states into two parts: $x = [x_1; x_2]$, where in practice $x_1 \in \mathbb{R}^{n_1}$ are states with slow dynamics and $x_2 \in \mathbb{R}^{n_2}$ are states with fast and stable dynamics. We consider the case where the trajectory and tracking feedback $u^{fb}$ are parameterized by only $x_1$.

**Definition 1**: *A locally Lipschitz continuous function $\gamma : \mathbb{R}^{n_1} \to \mathbb{R}^{n_2}$ such that $\gamma(0) = 0$ and satisfies*

$$\forall d \in \mathcal{D}, \eta_x(d) = [\eta_x^1(d); \eta_x^2(d)], \\ \eta_x^2(d) = \gamma(\eta_x^1(d)) \quad (44)$$

*is called an insertion map.*

The condition in (44) states that for any $d \in \mathcal{D}$, the insertion map maps the steady state of $x_1$ to the steady state of $x_2$. To extend the previous conclusion to cases where trajectories are parameterized with only $x_1$, we make the following assumptions.

**A-10**: $\exists \mathcal{S} \subseteq \mathbb{R}^n$, *compact, such that* $\forall d \in \mathcal{D}, \eta_x(d) \in \mathcal{S}$. *There exists CBF $b(x)$, such that.* $\forall x \notin \mathcal{S}, b(x) \leq 0$; $\forall d \in \mathcal{D}, b(\eta_x(d)) \geq 0$; $\forall d \in \mathcal{D}, \xi = [\xi_1; \xi_2] \in \mathcal{S}, \xi_2 = \gamma(\xi_1)$, *with $1 > c \geq 0$, there exists $u_\xi^d : [0, T_p] \to \mathbb{R}^m$ and a corresponding state trajectory, $\varphi_\xi^d : [0, T_p] \to \mathbb{R}^n$ satisfying*

$$V_d(\varphi_\xi^d(T_p) - \eta_x(d)) \leq cV_d(\varphi_\xi^d(0) - \eta_x(d)),$$

$$\frac{db}{dx}\big|_{\varphi_\xi^d(t)} \dot\varphi_\xi^d(t) + \kappa \varphi_\xi^d(t) \geq 0, \quad (45)$$

$$\lim_{[\xi_1, \gamma(\xi_1)] \to \eta_x(d)} \varphi_\xi^d(t) = \eta_x(d),$$

$$\lim_{[\xi_1, \gamma(\xi_1)] \to \eta_x(d)} u_\xi^d(t) \equiv \eta_u(d), \quad (46)$$

$$\varphi_{2\xi_1}^d(T_p) = \gamma(\varphi_{1\xi_1}^d(T_p)). \quad (47)$$

**A-11**: *There exists a feedback $u_1^{fb} : \mathbb{R}^{n_1} \times \mathbb{R}^{n_1} \to \mathbb{R}^m$ that $u_1^{fb}(x_1, \varphi_{1\xi_1}^d(\hat t))$ makes $\varphi_\xi^d$ uniformly locally exponentially stable, i.e. (43) is satisfied with $u(\hat t) = u_\xi^d(\hat t) + u_1^{fb}(x_1, \varphi_{1\xi_1}^d(\hat t))$.*

**Remark 9**: *The subscript $\varphi_{1\xi_1}^d$ means the desired trajectory of $x_1$ with initial condition $x_1(0) = \xi_1$, and $\varphi_{2\xi_1}^d$ means the desired trajectory of $x_2$, $\varphi_\xi^d = [\varphi_{1\xi_1}^d; \varphi_{2\xi_1}^d]$. A-11 is possible if the dynamic subsystem of $x_2$ is locally exponentially stable.*

**Theorem 2:** *Under A-5, A-6, A-7, A-10, and A-11, $\forall d \in \mathcal{D}$, $\forall \xi = [\xi_1; \gamma(\xi_1)] \in \{x \mid b(x) \geq 0\}$, the closed-loop system under CH feedback will stay inside $\{x \mid b(x) \geq 0\}$, and if $d$ stops changing after some $T \geq 0$, the state will converge to $\eta_x(d)$ exponentially.*

*Proof:* By A-11, the closed loop system exponentially converges to the CH desired trajectory. From A-10, by CBF condition, $\{x \mid b(x) \geq 0\}$ is invariant under the CH controller. When $d$ stops changing, the closed loop system exponentially converges to $\varphi_\xi^d$ and $V_d(x(nT_p) - \eta_x(d)) \leq c^{n-1} V_d(\xi - \eta_x(d))$, for $n = 1, 2, 3, \ldots$, and satisfies $x_2(nT_p) = \gamma(x_1(nT_p))$. So every time the desired trajectory is executed to the end, there exists $\varphi_{x_1(nT_p)}^d$ that follows the previous trajectory. By definition of the insertion map, (44) makes sure that when $x_1 \to \eta_x^1(d)$, $\gamma(x_1) \to \eta_x^2(d)$. By (46), $x(t)$ converges to $\eta_x(d)$ exponentially.

∎

**Remark 10**: *When the dynamics of $x_2$ is stable and fast, $y := x_2 - \varphi_{2\xi_1}$ converges to zero quickly, the influence of initial condition of $x_2$ is small enough to be neglected. Therefore, the CH can be parameterized only by $x_1$.*

Now consider a CH controller with trajectories generated with the procedure described in (28). A-5 and A-6 are trivially satisfied by the linear dynamics, where $\eta_x$ is defined such that it maps $r_d$ to the equilibrium point that renders $z = h(x) = 0$, which is unique. It can be shown that $\eta_x$ is Lipschitz continuous. We use the cost-to-go function $V$ of a Linear Quadratic Regulator (LQR) as the Lyapunov function by solving the Riccati equation. Since $V$ is quadratic, and the truck dynamic is linear, $V$ satisfies A-7 for all $r_d$. The CBF condition and Lyapunov condition in A-10 are enforced in the trajectory optimization by the last two constraints in (28). Pick $x_1 = [z \quad \dot z \quad \psi \quad -B_4 v_y + B_2 r \quad \psi_a]$, since $z$ and $\dot z$ are part of $x_1$, the closed loop dynamics is indeed stable under the PD control that only depends on $x_1$, which is the direct result of a stable zero dynamics, therefore satisfies the exponential stability condition in A-11. Note that the initial conditions in the training set are parameterized by $\Phi$, which is a full rank linear transformation of $x_1$ and $r_d$. By Theorem 2, the closed-loop system with CH feedback stays within $\{x \mid b(x) \geq 0\}$, and converges to $\eta_x(d)$ exponentially once $r_d$ stops changing.

### C. Smoothing of the desired trajectory

The smoothing of a Bezier curve is very simple. For an $m$-th order Bezier curve, the value for $0^{th}$ to $2^{nd}$ derivative at $s = 0$ are

$$\mathbf{B}(0) = \alpha_0,$$
$$\mathbf{B}'(0) = m\alpha_1 - m\alpha_0, \quad (48)$$
$$\mathbf{B}''(0) = m(m-1)(\alpha_2 + \alpha_0 - 2\alpha_1).$$

Solving for $\alpha_0$, $\alpha_1$ and $\alpha_2$:

$$\begin{aligned}
\alpha_0 &= h_{des}^0, \\
\alpha_1 &= \frac{\dot{h}_{des}^0}{m} + \alpha_0, \\
\alpha_2 &= \frac{\ddot{h}_{des}^0}{m(m-1)} + 2\alpha_1 - \alpha_0,
\end{aligned} \quad (49)$$

where $h_{des}^0$, $\dot{h}_{des}^0$ and $\ddot{h}_{des}^0$ are the value and derivatives of the desired trajectory before the update. The smoothing process requires that the Bezier order should be high enough so that the influence of the smoothing is limited to only the beginning of the curve. We choose the Bezier order to be 8.

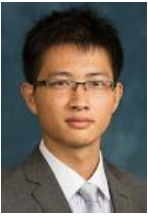

**Yuxiao Chen** is a Ph.D. candidate at the University of Michigan. He received his Bachelor's degree in Mechanical Engineering from Tsinghua University, Beijing, China in 2013. His research interests are control theory, cyber-physical system and particularly the application of vehicle control.

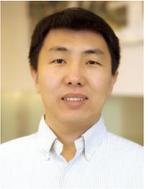

**Ayonga Hereid** received the Ph.D. in Mechanical Engineering from the Georgia Institute of Technology in 2016. He is currently a postdoctoral research fellow in the EECS department at the University of Michigan, Ann Arbor. His research interests lie at the intersection of nonlinear control and optimization theory, with a particular focus on developing elegant and principled control solutions for complex robotic systems, including bipedal robots and exoskeletons. He was the recipient of the Best Student Paper Award in 2014 from the ACM International Conference on Hybrid System: Computation and Control and was nominated as the Best Conference Paper Award Finalists in 2016 at the IEEE International Conference on Robotics and Automation.

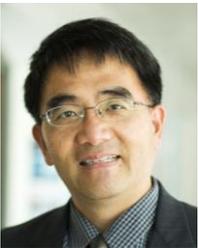

**Huei Peng** received his Ph.D. in Mechanical Engineering from the University of California, Berkeley in 1992. He is now a Professor at the Department of Mechanical Engineering at the University of Michigan. His research interests include adaptive control and optimal control, with emphasis on their applications to vehicular and transportation systems. His current research focuses include design and control of electrified vehicles, and connected/automated vehicles. Huei Peng has been an active member of the Society of Automotive Engineers (SAE) and the American Society of Mechanical Engineers. He is both an SAE fellow and an ASME Fellow. He received the National Science Foundation (NSF) Career award in 1998.

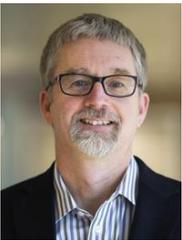

**Jessy W. Grizzle** received the Ph.D. in electrical engineering from The University of Texas at Austin in 1983. He is currently a Professor of Electrical Engineering and Computer Science at the University of Michigan, where he holds the titles of the Elmer Gilbert Distinguished University Professor and the Jerry and Carol Levin Professor of Engineering. He jointly holds sixteen patents dealing with emissions reduction in passenger vehicles through improved control system design. Professor Grizzle is a Fellow of the IEEE and IFAC. He received the Paper of the Year Award from the IEEE Vehicular Technology Society in 1993, the George S. Axelby Award in 2002, the Control Systems Technology Award in 2003, the Bode Prize in 2012 and the IEEE Transactions on Control Systems Technology Outstanding Paper Award in 2014. His work on bipedal locomotion has been the object of numerous plenary lectures and has been featured on CNN, ESPN, Discovery Channel, The Economist, Wired Magazine, Discover Magazine, Scientific American and Popular Mechanics.